# Magnetic Aharonov-Bohm effects and the quantum phase shift:

# A heuristic interpretation

Keith J. Kasunic

*Optical Systems Group LLC, Tucson, AZ 85728 USA*

Abstract

The shift in Aharanov-Bohm electron-interference fringe positions has been previously derived as resulting from phase differences induced by the magnetic vector potential **A**, without being clear on the physical mechanism behind it. In this paper, we show that the de Broglie wavelength of the electron wavefunction is changed locally by its interaction with the vector potential. The vector potential thus acts as a quantum "phase plate", changing the phase difference between interfering electron wavefunctions in a non-dispersive, gauge-invariant manner.

## I. BACKGROUND

In the well-known Aharonov-Bohm [1] and Ehrenberg-Siday [2] effects, the diffraction envelope of the electron wavepacket is not affected in a region of zero magnetic field **B**; instead, a phase difference – created by the magnetic vector potential **A** – changes the constructive- and destructive-interference fringe positions when electron wavefunctions are overlapped. As shown in Fig. 1, a typical experiment includes a solenoid placed beyond a two-slit diffraction geometry upon which the electron wavefunction is incident, and where the region outside the solenoid has zero magnetic field but finite vector potential.

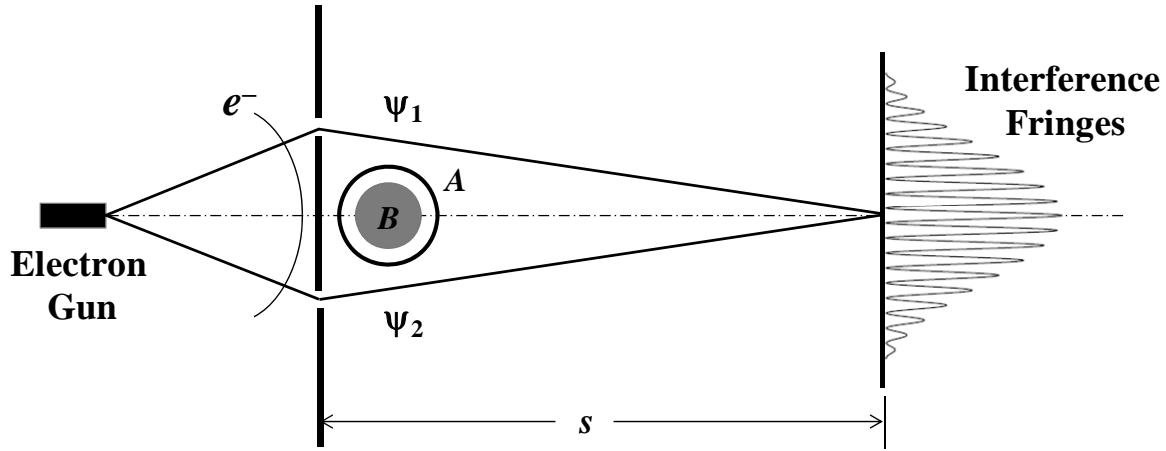

Figure 1 – Schematic of the Aharonov-Bohm (AB) and Ehrenberg-Siday (ES) two-slit interference experiments, where the magnetic field **B** is zero in the region outside the solenoid, yet the partial waves $\psi_1$ and $\psi_2$ for the electron $e^-$ recombine with different maxima and minima locations, depending on the vector potential **A**.

For small interference angles, the shift $\Delta y$ of the interference fringes due to the slit spacing (Fig. 2) – but not the diffraction envelope due to the slit width – can be determined from the geometric constraint on wavefunction interference, giving

$$\frac{\Delta y}{s} = \frac{\lambda_o}{d} \cdot \frac{e}{h} \Phi_B \tag{1}$$

for an initial electron wavelength $\lambda_o$, a slit spacing $d$, and a magnetic flux $\Phi_B$ through the solenoid, described in more detail below. For now, we note that the fringe shift depends on the magnetic flux, even though this quantity can be made to be zero in the free-space region outside the solenoid through which $\psi_1$ and $\psi_2$ propagate, even when there is a magnetic field inside the solenoid. As a result, there is no Lorentz force $e\mathbf{v} \times \mathbf{B}$ in the region outside the solenoid, and the electron wavepacket can therefore not move in response.

The vector potential $\mathbf{A}$, however – which can be related to the magnetic field via $\mathbf{B} = \nabla \times \mathbf{A}$ due to the absence of magnetic monopoles – is non-zero outside the solenoid. As a result, it has been proposed that $\mathbf{A}$ – conventionally thought of as a mathematical convenience rather than a physical variable – produces the experimentally-observed phase shifts [3]-[5].

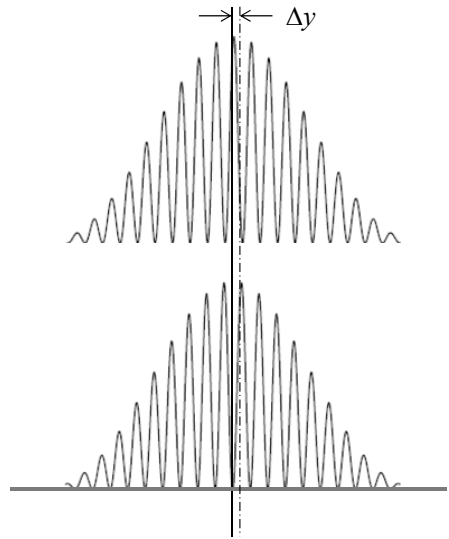

Figure 2 – Phase differences between $\psi_1$ and $\psi_2$ shift the constructive- and destructive-interference fringe positions, but do not affect the position of the overall diffraction envelope. With bright and dark fringes interchanged, a phase difference $\Delta\phi = e\Phi_B/\hbar = \pi$ radians between the top and bottom fringe patterns is illustrated in this figure.

Copyright © 2018 by Keith J. Kasunic.                                                                                          Page 3

Beyond mathematical convenience, James Clerk Maxwell and J. J. Thomson both considered the vector potential to carry field momentum. In the context of the AB effect, the phase shift has been described as a physical result of **A**-induced phase differences, without being clear on the specifics involved. Konopinski, for example, states that: "Thus a physical meaning of **A**, as a de Broglie 'phase shifter', was established." In addition, "…momentum exchanges between $M\mathbf{v}$ and the field momentum $q\mathbf{A}(\mathbf{r})/c$ played the essential role in shifting the phase." [6] Unfortunately, these statements were not further developed, thus leaving the reader to wonder as to the particulars behind the shift. The purpose of the paper is to propose a physical explanation as to how the AB phase shift occurs.

## II. ELECTRON WAVELENGTH and VECTOR POTENTIAL

As has been pointed out by many authors, it is not possible to analyze the quantum-mechanical properties of a system without the use of a canonical momentum that incorporates the vector potential **A**. As a result, the one-dimensional Hamiltonian for a non-relativistic particle such as a low-energy electron with charge $e$ depends on the mechanical momentum $\mathbf{p} = \mathbf{p_o} - e\mathbf{A}$ [1]

$$H = \frac{1}{2m}(\mathbf{p}_o - e\mathbf{A})^2 \qquad (2)$$

where $\mathbf{p_o}$ is the momentum of the electron before its interaction with the vector potential, and we have retained the 1D vector notation for $\mathbf{p}_o$ and **A**, as they may be pointed in either the same ("parallel") or opposing ("anti-parallel") directions. The gauge-invariant mechanical momentum thus determines the kinetic energy $T$ of the electron *during* the interaction

$$T = \frac{1}{2m}(\mathbf{p}_o - e\mathbf{A}) \cdot (\mathbf{p}_o - e\mathbf{A}) = \frac{p^2}{2m} = \frac{1}{2}mv^2 \qquad (3)$$

It is the mechanical momentum which determines the de Broglie wavelength of a moving particle. For an electron which has not interacted with the vector potential, the de Broglie wavelength $\lambda_o$ is given by

$$\lambda_o = \frac{h}{mv_o} = \frac{h}{p_o} \qquad (4)$$

where $mv_o = (2mE_o)^{1/2}$ for an electron with an initial energy $E_o$. From Eq. (3), the de Broglie wavelength $\lambda(\mathbf{A})$ is then changed by its local interaction with the vector potential

$$\lambda(\mathbf{A}) = \frac{h}{mv} = \frac{h}{|\mathbf{p}_o - e\mathbf{A}|} \qquad (5)$$

This equation illustrates that the de Broglie wavelength of a charged particle in a vector potential can be either longer or shorter than the non-interacting wavelength $\lambda_o$, depending on the vector sum determining the mechanical momentum $\mathbf{p} = \mathbf{p}_o - e\mathbf{A}$. As shown in Fig. 3, for example, we use the approximate method of Aharonov and Bohm by breaking the wavefunction down into partial waves: $\psi_1$ traversing the upper half the solenoid, and $\psi_2$ traversing the lower half. For a solenoidal magnetic field pointing out of the page, $\mathbf{p}_o$ and $\mathbf{A}$ are in the same direction on the lower half of the solenoid, thus reducing the mechanical momentum and increasing the de Broglie wavelength in this region of space. On the upper half of the solenoid, on the other hand, $\mathbf{p}_o$ and $\mathbf{A}$ are in opposing directions, thus increasing the mechanical momentum and decreasing the de Broglie wavelength. It is this change in de Broglie wavelength which changes the phase relationship between the electron's partial waves, thus laterally shifting the interference maxima and minima (fringes) which may occur in any region of space where the upper and lower wavefunctions are superimposed (interfered).

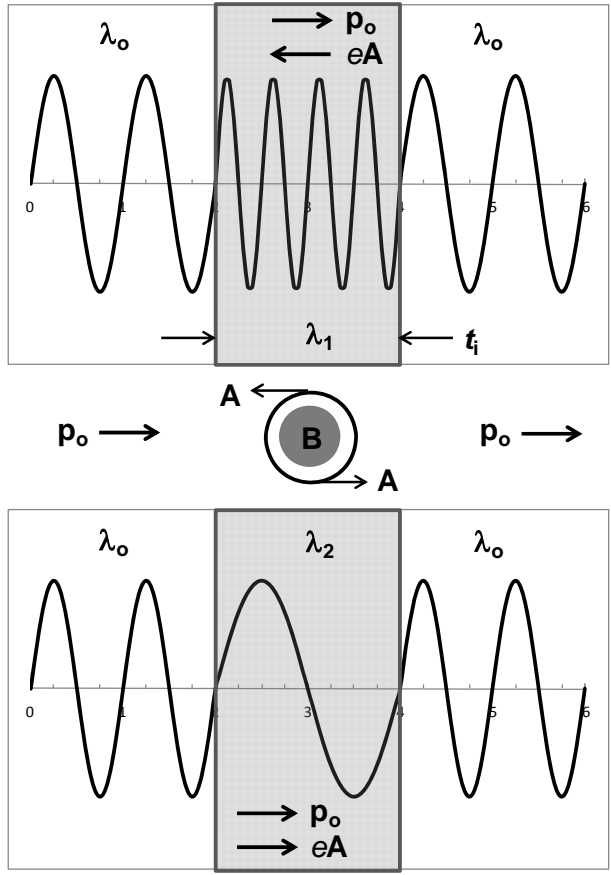

Figure 3 – The effects of the vector potential **A** on the de Broglie wavelength of an electron with an initial momentum $\mathbf{p_o}$ depend on the vector sum $\mathbf{p_o} - e\mathbf{A}$. For a magnetic field **B** pointing out of the page, this sum shortens $\lambda_1(A)$ for $\psi_1$ on the upper half of the solenoid, and lengthens $\lambda_2(A)$ for $\psi_2$ on the lower half.

To further develop this interpretation, we introduce the concept of a quantum refractive index (QRI) $n_q(\mathbf{A})$ – an idea briefly mentioned by Aharonov and Bohm in their 1959 paper [7]. By analogy with optical wave propagation – where the wavelength is shorter in a medium of higher optical refractive index (ORI) $n$ – we can define $n_q(\mathbf{A})$ by the ratio of either de Broglie wavelengths $\lambda_o/\lambda$ or group ("particle") velocities $v/v_o$

$$n_q(A) = \frac{\lambda_o}{\lambda(A)} = \frac{v}{v_o} = \frac{|\mathbf{p}_o - e\mathbf{A}|}{|\mathbf{p}_o|} \tag{6}$$

which reduces to a value of unity when $|\mathbf{A}| = 0$. Note that even though $n_q(\mathbf{A})$ is itself dispersive – i.e., it varies with initial electron momentum $p_o = \hbar k_o$ and de Broglie wavelength $\lambda_o$ – the phase *difference* across the solenoid is not.

To understand why, we model the $\mathbf{p}_o$-$\mathbf{A}$ interaction as a phase (or index) plate on both sides of the solenoid (Fig. 3), and write the quantum phase difference $\Delta\phi$ in an analogous manner with optical phase differences

$$\Delta\phi = \frac{2\pi}{\lambda_o} t_i \Delta n_q \tag{7}$$

where $t_i$ is the unknown thickness of each plate, and $\Delta n_q$ is obtained from Eq. (6)

$$\Delta n_q = n_{\psi_2}(A_2) - n_{\psi_1}(A_1) = \frac{p_o - eA_2}{p_o} - \frac{p_o - eA_1}{p_o} = \frac{e\Delta A}{p_o} = \frac{e}{h}\lambda_o \Delta A \tag{8}$$

for $p_o = h/\lambda_o$ and where the use of $\Delta\mathbf{A}$ makes $\Delta\phi$ gauge invariant. This invariance is a significant result, consistent with purely mathematical derivations of the phase shift [8] and illustrating in a simple equation the physical importance of differences in vector potential in determining these shifts.

The final expression for the phase shift is obtained by substituting the right-hand side of Eq. (8) in Eq. (7) and using $|\Delta\mathbf{A}| = 2A_\theta$ for a solenoidal field with a total interaction length $L_i = 2t_i$, giving

$$\Delta\phi = \frac{e}{\hbar} L_i A_\theta \tag{9}$$

Equation (9) is seen to be non-dispersive, as has been demonstrated experimentally by Caprez *et al.* by the absence of a net electromagnetic-force time delay on the electron wavepacket [9].

III. DISCUSSION

To compare our approach with previous theoretical and experimental results, we note that Aharonov and Bohm predicted a phase shift $\Delta\phi_{AB}$ [1]

$$\Delta\phi_{AB} = \frac{e}{\hbar}\oint A(r)\cdot ds = \frac{e}{\hbar}\iint B\cdot dS = \frac{e}{\hbar}\Phi_B \qquad (10)$$

for a circulatory line integral along a path length d**s**, equivalent (using **B** = $\nabla \times$ **A** and Stokes' theorem) to a flux integral over the solenoid area d**S**, with that same integral determining the magnetic flux $\Phi_B$. Using Eq. (10) to determine the maximum magnitude $|\mathbf{A}(R)| = A_\theta = \Phi_B/2\pi R$ oriented azimuthally at the outer radius $R$ of the solenoid, and comparing $\Delta\phi_{AB}$ with $\Delta\phi$ in Eq. (9), we find an interaction length $L_i = 2\pi R$, or the circumference of the solenoid.

This value of $L_i$ is fully consistent with Eq. (10) for an electron encircling the solenoid, thus confirming the validity of the de Broglie wavelength variations and resulting phase shifts found in Eq. (9). Also note that, as shown by the line integral in Eq. (10), the interaction length for $r > R$ outside the solenoid increases proportionally with the distance $r$ from the solenoid center, while the vector potential varies as $A_\theta = \Phi_B/2\pi r$, thus keeping the phase difference in Eq. (9) the same across the spatial extent of the wavefunction.

Experimentally, the changes in vector potential and QRI are typically quite small, resulting in extreme sensor sensitivity or requiring precise control over the solenoid's magnetic field. For example, for a phase shift $\Delta\phi_{AB} = 2\pi$ for the solenoid geometry shown in Fig. 1, Eq. (10) requires a magnetic flux $\Phi_B = h/e = 4.135 \times 10^{-15}$ Wb. Also from Eq. (10), the magnitude of the vector potential $A_\theta(R) = \Phi_B/2\pi R = 4.135 \times 10^{-15}$ Wb/($2\pi \times 5\times10^{-6}$ m) = $1.32 \times 10^{-10}$ Wb/m (or kg-m/sec$^2$-A) for a solenoid with an outer radius $R = 5$ μm. This gives a value of $eA = 2.11 \times 10^{-29}$ kg-m/sec to be used in Eq. (6) for determining $n_q(A)$. For an initial electron energy $E_o = eV_o = 1.602 \times 10^{-15}$ J ($V_o = 10$ kV), the incident electron momentum $p_o = (2mE_o)^{1/2} = 5.40 \times 10^{-23}$ kg-m/sec in Eq. (6) is larger than $eA$ by more than 6 orders of magnitude.

## IV. CONCLUSIONS

By changing the de Broglie wavelength of the electron wave function, the magnetic vector potential in the Aharonov-Bohm and Ehrenberg-Siday effects acts as a quantum "phase plate", changing in a non-dispersive, gauge-invariant manner the phase difference between $\psi_1$ and $\psi_2$ on the upper and lower halves of the solenoid. Physically, this is seen from Eq. (5) as a direct result of the vector potential adding (or subtracting) field momentum $e\mathbf{A}$ to (or from) the electron. We also note that this momentum transfer is local with $\mathbf{A}$, and there is no need to invoke a nonlocal interaction with $\mathbf{B}$ to describe the phase shift.

## Acknowledgements

This work has been supported by internal R&D funding at Optical Systems Group LLC.

## Endnotes and References

[1] Y. Aharonov and D. Bohm, "Significance of electromagnetic potentials in the quantum theory", Phys. Rev. **115** (3), 485–491 (1959).

[2] W. Ehrenberg and R. E. Siday, "The refractive index in electron optics and the principles of dynamics", Proc. Phys. Soc. (London), Sec. **B62**, 8–21 (1949).

[3] R. G. Chambers, "Shift of an electron interference pattern by enclosed magnetic flux", Phys. Rev. Lett. **5** (1), 3-5 (1960).

[4] See A. Tonomura *et al.*, "Observation of Aharonov-Bohm effect by electron holography", Phys. Rev. Lett. **48** (21), 1443-1446 (1982) and A. Tonomura *et al.*, "Evidence for Aharonov-Bohm effect with magnetic field completely shielded from electron wave", Phys. Rev. Lett. **56**, 792 (1986).

[5] Giulio Pozzi *et al.*, "Experimental realization of the Ehrenberg-Siday thought experiment", Appl. Phys. Lett. **108**, 083108 (2016).

[6] E. J. Konopinski, "What the electromagnetic vector potential describes", Am. J. Phys. **46** (5), 499-502 (1978).

[7] See Ref. 1, p. 490: "Indeed, the potentials play a role, in Schrodinger's equation, which is analogous to that of the index of refraction in optics."

[8] See, for example, R. P. Feynman, R. B. Leighton, and M. Sands, *The Feynman Lectures on Physics* (Addison-Wesley, 1965), Vol. 2, Chap. 15.

[9] Adam Caprez, Brett Barwick, and Herman Batelaan, "A macroscopic test of the Aharonov-Bohm effect", Phys. Rev. Lett. **99**, 210401 (2007).